\documentclass[preprint,secnumarabic,amssymb,showpacs, nobibnotes, aps, prd]{revtex4}
\usepackage{latexsym}
\usepackage{amsfonts,amsbsy}

\begin{document}
\title{de Sitter group and Einstein-Hilbert Lagrangian}

\author{Prasanta Mahato}
\email[email: ]{ pmahato@dataone.in}
\affiliation{ Department of Mathematics, Narasinha Dutt College\\
         Howrah, West Bengal, India 711 101 
         }
\begin{abstract}
 Axial vector torsion in the Einstein-Cartan space $U_{4}$ is considered here. By picking a particular
 term from the $SO(4,1)$ Pontryagin  density and then modifying it in a $SO(3,1)$ invariant way, we get a Lagrangian density with Lagrange multipliers. Then considering torsion and torsion-less
  connection as independent fields, it has been found that $\kappa$ and $\lambda$ of Einstein-Hilbert Lagrangian, appear as integration constants in such a way that $\kappa$ has been found to be linked with the topological Nieh-Yan density of $U_{4}$ space.
\end{abstract}
\maketitle
\section{Introduction}

In Riemannian manifold if metric and affine connection are treated
as two independent entities then the local geometry is endowed
with two independent tensors - curvature and torsion. Einstein
considered only curvature to be enough for the most economical and
successful theory of space-time. Cartan on the other hand
considered Einstein's view as a special case\cite{Deb79}. In the
recent background of abstract geometrical framework for a
consistent quantum theory of gravity the concept of torsion can be
easily welcomed.

  It is a natural belief that constants of nature have topological
  origin. In the well-known Einstein-Hilbert Lagrangian of gravity there are two constants -
  the gravitational constant and the cosmological constant. These constants are ad hoc in
  nature to justify classical gravitation. Moreover the constancy of the fundamental constants and, in particular, that of the gravitational constant has been questioned for a long time\cite{Dir37} and it led the early attempts to unify gravity with electromagnetism\cite{Kal21,Kle26}. Modern theories, like the string theory, link fundamental constants   with some extra dimensions and also predict variableness of these constants\cite{Pol94a}. Recent cosmological data\cite{Ton03} and its   analysis\cite{Gaz02} indicates a possible variation of the gravitational constant as it has been found that the peak luminosities of distant supernovae appear to be fainter than predicted by standard big bang cosmology. So it demands that in a generalized theory of gravity, like Scalar-Tensor theory, one should treat Newton's constant $G$ in  a different manner.

   It is well known\cite{Heh96} that torsion and curvature of any
  manifold are related to translation and rotation respectively. In particular torsion is more precisely related to broken translation gauge fields within the framework of nonlinear realization of the local space time group\cite{Tre00}. So to exploit both the symmetries of translation and rotation
  any gravitational Lagrangian   must contain torsion together with curvature.

  Today, gauge theory provides the theoretical base of all modern unification attempts in particle physics. In field theory gauge potentials become a standard tool for describing interactions with very different symmetries. And apparently the single gap in the modern gauge picture still remains : gauging the external or space-time symmetries of field and particles, that includes the gauge gravity also.

 Kibble\cite{Kib61} and Sciama\cite{Sci62} pointed out that the \textit{Poincar$\acute{e}$}  group, which is the semi-direct product of translation and Lorentz rotation, is the underlying gauge group of gravity and found the so-called Einstein-Cartan theory  where mass-energy of matter is related to the curvature and spin of matter is related to the torsion of space-time.

 Apart from being the most studied group\cite{Dre82}, the \textit{Poincar$\acute{e}$}  group has some remarkable properties\cite{Heh76}: \texttt{(1)} it is the basic local group of physics; \texttt{(2)} it classifies the elementary particles, giving them both spin and momentum; \texttt{(3)} it has a very clear relationship with space time. One  major drawback of  \textit{Poincar$\acute{e}$}  group is that it is a non-semi simple group which implies that there is no Lagrangian yielding its Yang-Mills equations\cite{Ald88}. There exists a general procedure\cite{Ald89} to check whether or not a set of field equations leads to a coherent theory, i.e. a theory that can be quantized. If we apply it to Yang-Mills equations for non-semi simple groups, we find that  they are never consistent. Here we see that though the \textit{Poincar$\acute{e}$} group is the classical group for relativistic kinematics it cannot be given a quantum version. Now by minimal addition of extra terms this inconsistent theory can be transformed to a good theory and  we   find a Lagrangian of a gauge theory for a semi-simple group, the de Sitter group, where vierbeins have been connected to the de Sitter boost parameters\cite{Ald88a}. In this way, the de Sitter gauge theory comes up as the corrected \textit{Poincar$\acute{e}$} gauge theory. Alternatively, there are other approaches where de Sitter group based Yang-Mills theories are shown to be producing either Ashtekar formulation of gravity\cite{Nie94} or Einstein-Cartan version of general relativity\cite{Can02}. One can also derive standard Einstein-Hilbert gravity from de Sitter gauge theory by choosing a particular Euler type four form as Lagrangian density\cite{Got90}.

 From geometrical point of view there is an important connection between the de Sitter group and the \textit{Poincar$\acute{e}$} group. It is a well known fact that the \textit{Poincar$\acute{e}$} group can be obtained from the  de Sitter group by an appropriate \textit{In$\ddot{o}$n$\ddot{u}$-Wigner} contraction\cite{Ino53,Dre85}. Here the contraction is achieved when  the de Sitter pseudo-radius $\mathcal{R}\rightarrow\infty$. By this contraction the  curvature of the de Sitter space $R\rightarrow 0$ where $R\propto \mathcal{R}$ ${}^{-2}$. We know that the de Sitter space is a solution of Einstein's equation for an empty space with cosmological constant $\Lambda =R/4$. Hence by the \textit{In$\ddot{o}$n$\ddot{u}$-Wigner} contraction, the de Sitter space reduces to Minkowski space which is a sourceless solution of Einstein's equation with a vanishing cosmological constant. Instead of taking $\mathcal{R}\rightarrow\infty$ if we take $\mathcal{R}\rightarrow$$ 0$ then we find that the de Sitter group reduces to the ``conformal"
 \textit{Poincar$\acute{e}$} group where the \textit{Poincar$\acute{e}$} translation is replaced by the special conformal transformation  and the de Sitter space reduces to a cone-space whose geometry is gravitationally related to an infinite cosmological constant\cite{Ald98}.

 Recently a gravitational Lagrangian has been been proposed\cite{Mah02a} where a
 Lorentz invariant part of the de Sitter Pontryagin density has been treated as the Einstein-Hilbert Lagrangian. In this paper we shall try
  to establish the constancy of this gravitational constant under the back ground where
  the cosmological constant is the only source of gravitation.


 \section{Topological densities and gravity  Lagrangian  }
 Cartan's structural equations for a Riemann-Cartan space-time $U_{4}$ are given by\cite{Car24}
 \begin{eqnarray}T^{a}&=& de^{a}+\omega^{a}{}_{b}\wedge e^{b}\label{eqn:ab}\\
 R^{a}{}_{b}&=&d\omega^{a}{}_{b}+\omega^{a}{}_{c}\wedge \omega^{c}{}_{b},\label{eqn:ac}\end{eqnarray}
here $\omega^{a}{}_{b}$ and e$^{a}$ represent the spin connection
and the local frames respectively.

In $U_{4}$ there exists  two invariant closed four forms. One is the well
known Pontryagin\cite{Che74} density \textit{P} and the other is
the less known Nieh-Yan(NY)\cite{Nie82} density \textit{N} given by
\begin{eqnarray} \textit{P}&=& R^{a}{}_{b}\wedge R^{b}{}_{a}\hspace{1mm},\label{eqn:ad}\\ \textit{N}&=& d(e_{a}\wedge T^{a})\nonumber \\
&=&T^{a}\wedge T_{a}- R_{ab}\wedge e^{a}\wedge
e^{b}.\label{eqn:af}\end{eqnarray}

 The minimal Lagrangian density of a spin-$\frac{1}{2}$ field $\psi$ with an external gravitational
 field with torsion is given
by\cite{Mie01}\begin{eqnarray}L_{D}&=&\frac{i}{2}\{\overline{\psi}{}^* \gamma \wedge D\psi+\overline{D\psi}\wedge{}^*\gamma\psi\}+{}^*m\overline{\psi}\psi\nonumber\\&{}&-\frac{1}{4}A\wedge\overline{\psi}\gamma_5{}^*\gamma\psi,\label{eqn:ayx}\end{eqnarray}where $\gamma=\gamma^a e_a$,  $D$ is the torsion-free exterior derivative,
$A$ is the axial vector part of the   torsion one form and ${}^*$ is the Hodge duality operator. Therefore,
considering the source in the matter Lagrangian, we can simply
assume that the torsion is given by an axial vector only.

 In presence of
  axial vector  torsion, one naturally gets the Nieh-Yan
density from  (\ref{eqn:af})
\begin{eqnarray} N&=&-R_{ab}\wedge e^{a}\wedge e^{b}=-{}^* N\eta\hspace{2 mm},\label{eqn:xaa}\\
 \mbox{where} \hspace{2 mm}\eta&:=&\frac{1}{24}\epsilon_{abcd}e^{a}\wedge e^b\wedge e^c\wedge e^d\end{eqnarray}is the invariant volume element.  It follows that    ${}^*N$, the Hodge dual of $N$, is a scalar density of dimension $(length)^{-2}$.

 It has been shown in an earlier paper\cite{Gho00}, when a `direction vector' (vortex line) is attached with a space-time point $x_\mu$, this eventually leads to the $SL(2,C)$ gauge theory of gravitation including torsion. The space time manifold now corresponds to the de Sitter space $M^{4,1}$. In this framework $SO(4,1)$ Pontryagin density can be written as  \cite{Cha97,Mah02a}
\begin{eqnarray}R^{A}{}_{B}\wedge R^{B}{}_{A}&=&R^{a}{}_{b}\wedge R^{b}{}_{a}+
\frac{2}{l^{2}}N\nonumber\\&=
&\tilde{R}^{a}{}_{b}\wedge\tilde{R}^{b}{}_{a}-\frac{1}{3}RN+
\frac{2}{l^{2}}N,\label{eqn:xb}\\\mbox{where}\hspace{2
mm}\tilde{R}^{ab}&:=&R^{ab}-\frac{1}{12}{}^*N\eta^{ab},\label{eqn:xb1}\\\eta_{ab}&:=&\frac{1}{2!}\epsilon_{abcd}e^c\wedge e^d,\\\mbox{such that}\hspace{2
mm}R&=&\tilde{R}
=\frac{1}{2}{}^*(\tilde{R}^{ab}\wedge \eta_{ab}),\label{eqn:xb2}\\\mbox{and}\hspace{5mm}\tilde{R}^{ab}\wedge e_a\wedge e_b&=&(R^{ab}-\frac{1}{12}{}^*N\eta^{ab})\wedge e_a\wedge e_b\nonumber\\&=&0,\label{eqn:xb3}\end{eqnarray} $l$  is a
fundamental length constant and $A,B=0,1,..,4.$
Usually,  when Lorentz group is embedded into the de Sitter group, $l$ is called   the radius of the universe and is related to the cosmological constant\cite{Cha97}. The above decomposition (\ref{eqn:xb}), (\ref{eqn:xb1}) is always possible in $U_4$ provided, the divergence four form of the axial-vector torsion, $N\neq 0$. This serves no restriction on the gauge   field part of the $SO(3,1)$ connection one form $\omega^{ab}$ in the background of a broken $SO(4,1)$ gauge theory. In other word the above breakup guarantees that, in the tangent space, only $SO(3,1)$ symmetry is preserved and $SO(4,1)$ symmetry is broken. The two parts $\tilde{R}^{ab}$ and $\frac{1}{12}{}^*N\eta^{ab}$ of $R^{ab}$ are two independent covariant two forms under $SO(3,1)$ rotation in the tangent space. From   (\ref{eqn:xb2}) and (\ref{eqn:xb3}) we see that $\tilde{R}^{ab}$ and $\frac{1}{12}{}^*N\eta^{ab}$ are respectively  connected to $R$ and $N$. Hence by the above decomposition we are separating two irreducible parts of the curvature w. r. t. the one form $e^a$   and consequently neither any Bianchi identity is violated nor it leads to any constraint on $N$.

Recently, it has been shown that\cite{Mah02a}, in $U_{4}$ space
one can locally consider a particular term from this $SO(4,1)$
Pontryagin density as the gravitational Lagrangian\cite{note1}, given by
\begin{eqnarray}\mathcal{L}\mbox{${}_{0}=   -{}^*N R\eta $}.\end{eqnarray}
  This Lagrangian looks like the Einstein-Hilbert Lagrangian,
  \begin{eqnarray}\mathcal{L}\mbox{${}_{EH}=\frac{1}{\kappa}R\eta,$}\end{eqnarray}
provided ${}^*N=\frac{1}{\kappa}, $ where $\kappa$ is
Einstein's gravitational constant, and   the torsional
contribution disappears from the scalar curvature $R$.
It is a well-known result\cite{Sha01} in $U_{4}$ space, in the
case of axial vector torsion, that
\begin{eqnarray}R=R_{E}+\frac{1}{4}{}^*(T\wedge{}^*T),\label{eqn:abc}\end{eqnarray}
where $R_{E}$ represents scalar curvature when the connection is
without torsion and $T$ is the torsion three form representing axial vector torsion $A$. Now if we assume the axial vector $A$ to be
a null vector then (\ref{eqn:abc}) reduces to
\begin{eqnarray}R=R_{E}.\label{eqn:abcd}\end{eqnarray}
This can be
guaranteed by introducing a $SO(3,1)$ invariant Lagrangian density
$\mathcal{L}\mbox{$_{1}$}$, given by
\begin{eqnarray}\mathcal{L}\mbox{$_{1}= \zeta{}^*(T\wedge{}^*T)(T\wedge{}^*T),$}\end{eqnarray}
where $\zeta$ is a dimensionless Lagrange multiplier.    So far $SO(3,1)$
invariance is concerned, torsion can be separated from the
connection  as torsional part of the $SO(3,1)$ connection
transforms like a tensor i.e. when the local frame also transforms like a
$SO(3,1)$ valued one form (not connection one form) in a broken $SO(4,1)$ gauge theory. In this
direction it is important to define a torsion-free exterior
differentiation through a field equation involving the connection
and the local frame only. So  we introduce another Lagrangian
density $\mathcal{L}\mbox{${}_{2}$}$, given by,
\begin{eqnarray}\mathcal{L}\mbox{$_{2}$}=\mbox{${}^*(b_a\wedge
\nabla e^{a})(b_a\wedge
\nabla e^{a})$} ,\end{eqnarray} where   $\nabla$
represents exterior differentiation with respect to a $SO(3,1)$
connection one form $\bar{\omega}^{ab}$ and $b_{  a}$ is a two form with
one internal index and of dimension $(length)^{-1}$. If we treat $b_{a}$ as Lagrange multiplier
then it ensures that $\nabla$ represents torsion-free exterior
differentiation. By this way torsion has become decoupled from the connection part of the theory. It has become independent of the one form $e^a$, in particular, owing to its fundamental existence as a metric independent tensor in the affine connection in $U_4$, we treat here the three form $T=e^a\wedge T_a$ as more fundamental than the one form $T^{ab}=\omega^{ab}-\bar{\omega}^{ab}$. Without any ambiguity and for future consistency,
we can consider a connection independent cosmological density,
given by
\begin{eqnarray}\mathcal{L}\mbox{$_{3}=-\Lambda \eta$} \hspace{2
mm},\end{eqnarray}where $\Lambda$ is a constant whose value is to
be ascertained later on.  Now we are in a position to define the
total gravitational Lagrangian density in empty space as
\begin{eqnarray}\mathcal{L}\mbox{$_{G}$}&=&\mathcal{L}\mbox{$_{0}+$}\mathcal{L}\mbox{$_{1}+$}\mathcal{L}\mbox{$_{2}+$}\mathcal{L}\mbox{$_{3}$}\hspace{2 mm},\label{eqn:abl}\\
&=&-{}^*N R\eta +\zeta{}^*(T\wedge{}^*T)(T\wedge{}^*T)\nonumber\\
&{}& +\hspace{1 mm}{}^*(b_a\wedge
\nabla e^{a})(b_b\wedge
\nabla e^{b}) -\Lambda \eta \hspace{2 mm},\nonumber\end{eqnarray}where $N=dT$ and $R\eta=-\frac{1}{2} (d\bar{\omega}^{ab}+\bar{\omega}^a{}_f\wedge\bar{\omega}^{fb})\wedge \eta_{ab}$.  To
start with this Lagrangian we have altogether 69 independent
components of the field variables $e^{a}$, $T$,
 $\bar{\omega}^{ab}$,   $b^{a}$ and $\zeta$.
 \section{Euler-Lagrange equations and  gravitational constant}
 The Lagrangian $\mathcal{L}\mbox{$_{G}$}$, which is defined in the previous section,
  is  only Lorentz invariant  under rotation in the tangent space where  de Sitter boosts
   are not permitted. As a consequence $T$ can be treated independently of $e^a$ and
   $\bar{\omega}^{ab}$. Then following reference \cite{Heh95}, we independently vary
   $e^{a}$, $\nabla e^{a}$, $T$, $dT$, $R^{ab}$,
    $b^{a}$ and $\zeta$ and find

\begin{eqnarray}
    \delta  \mathcal{L}\mbox{$_{G}$}&=&\delta e^a\wedge \frac{\partial
    \mathcal{L}\mbox{$_{G}$}}{\partial e^a}+\delta \nabla e^a\wedge \frac{\partial
    \mathcal{L}\mbox{$_{G}$}}{\partial \nabla e^a}+\delta T\wedge\frac{\partial  \mathcal{L}
    \mbox{$_{G}$}}{\partial T}\nonumber\\&{}&+\delta dT \frac{\partial  \mathcal{L}\mbox{$_{G}$}}
    {\partial dT} +\delta R^{ab}\wedge \frac{\partial  \mathcal{L}
    \mbox{$_{G}$}}{\partial R^{ab}}+\delta b^a\wedge \frac{\partial  \mathcal{L}
    \mbox{$_{G}$}}{\partial b^a}\nonumber\\&{}&+\delta \zeta \frac{\partial  \mathcal{L}
    \mbox{$_{G}$}}{\partial \zeta}\\&=&\delta e^a\wedge( \frac{\partial
    \mathcal{L}\mbox{$_{G}$}}{\partial e^a}+ \nabla\frac{\partial
    \mathcal{L}\mbox{$_{G}$}}{\partial \nabla e^a})+\delta T\wedge(\frac{\partial
    \mathcal{L}\mbox{$_{G}$}}{\partial T}+ d \frac{\partial  \mathcal{L}\mbox{$_{G}$}}
    {\partial dT})\nonumber\\&{}&+\delta \bar{\omega}^{ab}\wedge(\nabla \frac{\partial
     \mathcal{L}\mbox{$_{G}$}}{\partial R^{ab}}+ \frac{\partial
    \mathcal{L}\mbox{$_{G}$}}{\partial \nabla e^a}\wedge e_b)+\delta b^a\wedge \frac{\partial
     \mathcal{L}\mbox{$_{G}$}}{\partial b^a}\nonumber\\&{}&+\delta \zeta \frac{\partial
     \mathcal{L}\mbox{$_{G}$}}{\partial \zeta}+d(\delta e^a\wedge
     \frac{\partial  \mathcal{L}\mbox{$_{G}$}}{\partial \nabla e^a}+\delta T \frac{\partial
      \mathcal{L}\mbox{$_{G}$}}{\partial dT}\nonumber\\&{}&+\delta \bar{\omega}^{ab}\wedge \frac{\partial
      \mathcal{L}\mbox{$_{G}$}}{\partial R^{ab}})\label{eqn:abc0}
\end{eqnarray}
Using the form of the Lagrangian $\mathcal{L}\mbox{$_{G}$}$, given in (\ref{eqn:abl}),
we get
\begin{eqnarray}
    \frac{\partial  \mathcal{L}\mbox{$_{G}$}}{\partial e^a}&=&-{}^*N(2\textbf{R}_a-R\eta_a)
    +\zeta\tau(-12\tau^b{}_a\eta_b+\tau\eta_a)\nonumber\\&{}&+2{}^*(b_b\wedge
\nabla e^{b})^2\eta_a-\Lambda\eta_a\label{eqn:abc1} \\  \frac{\partial  \mathcal{L}
\mbox{$_{G}$}}{\partial (\nabla e^a)}&=&2{}^*(b_a\wedge \nabla e^{a})b_a\label{eqn:abc2}
\\\frac{\partial  \mathcal{L}\mbox{$_{G}$}}{\partial T}&=&4\zeta{}^*(T\wedge{}^*T){}^*T\label{eqn:abc3} \\\frac{\partial  \mathcal{L}\mbox{$_{G}$}}{\partial (dT)}&=&R\label{eqn:abc4}\\\frac{\partial  \mathcal{L}\mbox{$_{G}$}}{\partial R^{ab}}&=&\frac{1}{4}{}^*N\epsilon_{abcd}e^c\wedge e^d=\frac{1}{2}{}^*N\eta_{ab}\label{eqn:abc5}\\ \frac{\partial  \mathcal{L}\mbox{$_{G}$}}{\partial b^a}&=&2{}^*(b_b\wedge
\nabla e^{b})
\nabla e_{a}\label{eqn:abc6}\\\frac{\partial  \mathcal{L}\mbox{$_{G}$}}{\partial \zeta}&=&{}^*(T\wedge{}^*T)(T\wedge{}^*T)\label{eqn:abc7}
\end{eqnarray}
  Where
\begin{eqnarray}
 \textbf{R}_a&:=&\frac{1}{2}\frac{\partial (R\eta)}{\partial e^a}=\frac{1}{4}\epsilon_{abcd}R^{bc}\wedge  e^d=-G^b{}_a\eta_b\label{eqn:abc60}\\G^b{}_a&:=&R^b{}_a-\frac{1}{2}R\delta^b{}_a\\\eta_a&:=&\frac{\partial \eta}{\partial e^a}=\frac{1}{3!}\epsilon_{abcd}e^b\wedge e^c\wedge e^d\\\tau^a{}_b&:=&T^{a\mu\nu}T_{b\mu\nu}\hspace{5mm}\mbox{and}\hspace{5mm}\tau:=\tau^a{}_a
\end{eqnarray}From above, Euler-Lagrange equations for $\zeta$ and $b_a$ give us
\begin{eqnarray}
    T\wedge{}^*T&=&0\label{eqn:abc8}\\
\nabla e_{a}&=&0\label{eqn:abc9}
\end{eqnarray}i.e. $T$ corresponds to a null axial vector and $\nabla$ is torsion free. Using this result in (\ref{eqn:abc1}), (\ref{eqn:abc2}) and (\ref{eqn:abc3}) we get
\begin{eqnarray}
    \frac{\partial  \mathcal{L}\mbox{$_{G}$}}{\partial e^a}&=&-{}^*N(2\textbf{R}_a-R\eta_a)-\Lambda\eta_a\label{eqn:abc10}\\    \frac{\partial  \mathcal{L}\mbox{$_{G}$}}{\partial (\nabla e^a)}&=&0\label{eqn:abc11}\\\frac{\partial  \mathcal{L}\mbox{$_{G}$}}{\partial T}&=&0\label{eqn:abc12}
\end{eqnarray}Using these results in equations from (\ref{eqn:abc0}) to (\ref{eqn:abc5}), we get Euler-Lagrange equations of $e^a$, $T$ and $\bar{\omega}^{ab}$, given by
\begin{eqnarray}
    {}^*N(2\textbf{R}_a-R\eta_a)+\Lambda\eta_a&=&0\label{eqn:abc13}\\dR&=&0\label{eqn:abc14}\\\nabla({}^*N\eta_{ab})&=&0\label{eqn:abc15}
\end{eqnarray}Using (\ref{eqn:abc9}), the last equation yield
\begin{eqnarray}
    d{}^*N=0\label{eqn:abc16}
\end{eqnarray}From  equations (\ref{eqn:abc14}) and (\ref{eqn:abc16}) we can write
\begin{eqnarray}
{}^*N=\frac{1}{\kappa}\hspace{4mm}\mbox{and}\hspace{4mm}R=2\lambda\hspace{4mm}\mbox{(say).}\label{eqn:abc17}
\end{eqnarray}
Using (\ref{eqn:abc17}) in (\ref{eqn:abc13}) and then using (\ref{eqn:abc60}) we get
\begin{eqnarray}
    G^b{}_a=-\frac{1}{2}\lambda\delta^b{}_a, \label{eqn:abc18}
\end{eqnarray}where, for consistency, $\lambda=\kappa\Lambda$. This last equation is the   Einstein's equation of gravity in the presence of the cosmological constant $\lambda$ corresponding to the Einstein-Hilbert Lagrangian
\begin{eqnarray}\mathcal{L}\mbox{$_{EH}^{CC}=\frac{1}{\kappa}(R-\lambda)\eta$}.\label{eqn:abc19}
\end{eqnarray}

 We know that, though torsion one form $T^{ab}=\omega^{ab}-\bar{\omega}^{ab}$ is a part of
the $SO(3,1)$ connection, it
 does not transform like a connection  form under $SO(3,1)$   rotation in
 the tangent space  and thus it imparts no constraint on the gauge degree of freedom of the
   Lagrangian. By this way the role of torsion in the underlying manifold has become multiplicative
   rather than additive one and the quadratic first part of the  Lagrangian $\mathcal{L}\mbox{$_{G}$}$ looks like
    $torsion \otimes curvature$\cite{note2}. In other words - the additive torsion is decoupled from the
   theory but not the multiplicative one. This indicates that torsion is uniformly nonzero
   everywhere. In the geometrical sense, this implies that
   micro local space-time is such that at every point there is a
   direction vector (vortex line) attached to it. This effectively
   corresponds to the non commutative geometry having the manifold
   $M_{4}\times Z_{2}$ where the discrete space $Z_{2}$ is just
   not the two point space\cite{Con94} but appears as an attached direction vector. This has direct relevance in the quantization of a fermion where the discrete space appears as the internal space of a particle\cite{Gho00}. This becomes relevant if we
   consider that fermions are the basic building units of matter.
   The existence of a globally defined null vector field ${}^*T$
   with non-zero divergence then corresponds to the axial vector
   current leading to chiral anomaly. Now, in the background of the minimal action of a   spinor field given in (\ref{eqn:ayx}) and even in massless case,   there is a divergent   contribution of torsion to chiral anomaly given by\cite{Cha97}
\begin{eqnarray}
    d \left\langle \begin{array}{c}j_5\end{array}\right\rangle=\textit{A}(x)
\end{eqnarray}where \[j_5=\bar{\psi}{}^*\gamma\gamma_5\psi,\hspace{2mm}\gamma=\gamma^ae_a\] and \[\textit{A}(x)=2\sum_{n}\eta\psi_n^\dag\gamma_5\psi_n.\] Then under standard regularization by the square of the Dirac operator  in Einstein-Cartan space\cite{Cha97}
\begin{eqnarray}
\textit{A}&=&\lim_{M\rightarrow\infty}\frac{1}{8\pi^2}[R^{ab}\wedge
R_{ab}+2M^2(T_a\wedge T^a\nonumber\\&{}&-R_{ab}\wedge e^a\wedge
e^b)+O(M^{-2})]
\end{eqnarray}
As chiral anomaly appears
   as the quantum mechanical symmetry breaking, torsion in this
   sense represents the quantum effect\cite{Ban91}. Mielke et al have questioned the contribution of the NY term to the chiral anomaly as well as its non triviality after  renormalization\cite{Mie99,Mie99a,Kre01}. Contribution of the NY term to chiral anomaly has been confirmed by Obukhov et al\cite{Obu97} and in an independent analysis\cite{Cha98a}, computing the index of the Dirac operator on a four dimensional compact manifold, it has been shown that the integral of  the NY term is necessarily an integer,  it is the difference of two Chern classes $SO(5)$ and $SO(4)$ and therefore being topological $N$ is non-renormalizable. In our present analysis this bears an important implication as we see from (\ref{eqn:abc17}) that topological $N$ globally defines the gravitational constant, at least in the case where the cosmological constant is the only source of gravity, by the equation
\begin{eqnarray}
    N=-\frac{1}{\kappa}\eta.
\end{eqnarray}
It is important to note that, in some other approach\cite{Mie03},     some multiplicative $torsion\otimes curvature$ terms also appear in $3$D gravity but they seem to be devoid of any topological interpretation. Such terms do not appear in $4$D.

Hence (\ref{eqn:abc18})  implies that our starting
Lagrangian $\mathcal{L}\mbox{$_{G}$}$ is equivalent to the Einstein-Hilbert
Lagrangian in vacua in presence of a cosmological constant where
the two constants $\kappa$ and $\lambda$ are constants of
integration but also,  by (\ref{eqn:abc17}) ,
$\lambda$ is half of the $SO(3,1)$ scalar curvature and  the topological Nieh-Yan density is $(-\frac{1}{\kappa})$ times the invariant volume element. Also the form of the starting Lagrangian $\mathcal{L}\mbox{$_{G}$}$ implies that constancy of the gravitational constant depends upon the fact that  the source term in $\mathcal{L}\mbox{$_{G}$}$ is independent of the $SO(3,1)$ gauge connection. This suggests that in a gravity theory with $SO(3,1)$ connection, torsion,   local frame and matter,   if we want  gravitational constant   to be an on-shell constant, then the material source term should be independent of the $SO(3,1)$ connection.

\section{Discussion}
Recent cosmological evidence\cite{Fil03,Ton03} suggests that
cosmological constant  seems to be present evermore in the
cosmological data. Theoretically, cosmological constant appears
when one considers a four dimensional manifold that is due to
compactification\cite{note3} of a five dimensional manifold with
the signature of a (anti)de Sitter space time\cite{Pad03}. This
implies that in the local tangent space the gauge group structure
is either $SO(4,1)$ or $SO(3,2)$. To keep Lorentz invariance
intact  (anti)de Sitter boost is forbidden in the tangent space.
So it is justified to consider the Lagrangian as a particular
$SO(3,1)$ invariant part of the full $SO(4,1)$   Pontryagin
density.

 It is important to  note  that, in our present formalism, the only
 assumption is that the torsion is represented by a null axial
 vector   and the corresponding Lagrangian is a particular term
 of the $SO(4,1)$ Pontryagin density in such a way that the $SO(3,1)$
 invariance of the theory is maintained. The presence of the null axial vector at each space time point suggests that the space time manifold is characterized by the presence of a `direction vector'(vortex line) attached to each point which is the source of torsion. It may be remarked that the    degrees of freedom
 of this theory is minimally extended from that of
 Einstein-Hilbert theory with torsion contributing to the
 additional degree. As a result $\kappa$ and $\lambda$ have got
 their definite geometrical meaning in $U_{4}$ space in
 comparison to  their standard meaning of being simply constants such that, in empty space,   $\lambda$ is half of the $SO(3,1)$ scalar curvature and ($-\frac{1}{\kappa}$)
is   the proportionality constant between the topological Nieh-Yan density
and the invariant volume four form $\eta$.
  Moreover, being constants of integration, $\kappa$ and $\lambda$ might have
  got their fixed values in the Early Universe when the bulk
  matter was created.

In a recent paper\cite{Mah02} it has been shown that, in the
gravity without metric formalism of gravity, when one performs a
particular canonical transformation of the field variables,
CP-violating $\theta$-term appears in the Lagrangian together with
the cosmological term. This supports the finding of this note when
we consider that the torsion, being an axial vector, has a certain
role to play in CP-violation. Indeed,   the topological $\theta$-term
 of \textquoteleft gravity without metric formalism' is linked with
 the topological Nieh-Yan density of U$_{4}$
 geometry.   CP-violation  or non-violation by topological terms has   been discussed also in reference\cite{Mie01}. Link of torsion   with CP-violation can  be found also in   $SL(2,C)$ gauge approach of gravity\cite{Ban95}. Thus arrow of time may play a significant role in the geometrical origin of torsion and hence of the gravitational constant.
 \section*{ACKNOWLEDGEMENT}

   I wish to thank Prof. Pratul Bandyopadhyay,
   Indian Statistical Institute, Kolkata,
    for his valuable remarks and fruitful
    suggestions on this
           problem.



\begin{thebibliography}{10}

\bibitem{Deb79}
R.~Debever(Ed.).
\newblock {\em Elie Cartan-Albert Einstein Letters on Absolute Parallelism}.
 Princeton University Press, Princeton, (1979).

\bibitem{Dir37}
P.~A.~M. Dirac.
 Nature \newblock {\bf 139}, 323, (1937).

\bibitem{Kal21}
T.~Kaluza.
  Preuss. Akad. Wiss. K \newblock {\bf 1}, 966, (1921).

\bibitem{Kle26}
O.~Klein.
 Z. Physik \newblock {\bf 37}, 895, (1926).

\bibitem{Pol94a}
T.~Damour and A.~M. Polyakov.
  Nucl. Phys. B \newblock {\bf 423}, 532, (1994).

\bibitem{Ton03}
J.~N.~Tonry et~al.
 Astron. J. \newblock {\bf 594}, 1, (2003).

\bibitem{Gaz02}
E.~Gaztanaga et~al.
 Phy. Rev. D \newblock {\bf 65}, 023506, (2002).

\bibitem{Heh96}
F.~W. Hehl.
\newblock {\em Proc. of the 14th course of the School of Gravitation and
  Cosmology and Gravitation on Quantum Gravity, held at Erice, Italy May
  1995}(Eds. P. G. Bergman, V. de Sabbata and H. J. Treder), World Scientific,
  Singapore, (1996).

\bibitem{Tre00}
R.~Tresguerres and E.~W. Mielke.
  Phys. Rev. D \newblock {\bf 62}, 44004, (2000).

\bibitem{Kib61}
T.~W.~B. Kibble.
  J. Math. Phys. \newblock {\bf 2}, 212, (1961).

\bibitem{Sci62}
D.~W. Sciama.
\newblock {\em ``On the analogy between charge and spin in general relativity"
  : in Recent Developments in General Relativity}, (Pergamon + PWN), Oxford,
  page 415, (1962).

\bibitem{Dre82}
W.~Drechsler.
  Ann. Inst. Henri Poincar\'{e} \newblock {\bf 37}, 155, (1982).

\bibitem{Heh76}
{G. W. Hehl, P. von der Heyde} and G.~D. Kerlick.
  Rev. Mod. Phys. \newblock {\bf 48}, 393, (1976).

\bibitem{Ald88}
R.~Aldrovandi and R.~A. Kraenkel.
  J. Phys. A \newblock {\bf 21}, 1329, (1988).

\bibitem{Ald89}
R.~Aldrovandi and R.~A. Kraenkel.
  J. Math. Phys. \newblock {\bf 30}, 1966, (1989).

\bibitem{Ald88a}
R.~Aldrovandi and J.~G. Pereira.
  J. Math. Phys. \newblock {\bf 29}, 1472, (1988).

\bibitem{Nie94}
{J. A. Nieto, O. Obreg$\acute{\mbox{o}}$n} and J.~Socorro.
  Phys. Rev. D \newblock {\bf 50}, 3583, (1994).

\bibitem{Can02}
M.~Botta Cantcheff.
  Gen. Rel. Grav. \newblock {\bf 34}, 1781, (2002).

\bibitem{Got90}
S.~Gotzes and A.~C. Hirshfeld.
  Ann. Phys. (NY) \newblock {\bf 203}, 410, (1990).

\bibitem{Ino53}
E.~In$\ddot{\mbox{o}}$n$\ddot{\mbox{u}}$ and E.~P. Wigner.
  Proc. Natl. Acad. Scien. \newblock {\bf 39}, 510, (1953).

\bibitem{Dre85}
W.~Drechsler.
  J. Math. Phys. \newblock {\bf 26}, 41, (1985).

\bibitem{Ald98}
R.~Aldrovandi and J.~G. Pereira.
\newblock {\em A contribution to Conference on Topics in Theoretical Physics
  II: Festschrift for A. H. Zimmerman}, Sao Paulo, Brazil, 20 Nov, 1998,
  \newblock {\bf 16}, 495, (1999).

\bibitem{Mah02a}
P.~Mahato.
  Mod. Phys. Lett. A \newblock {\bf 17}, 1991, (2002).



\bibitem{Car24}
E.~Cartan.
  Ann. Ec. Norm. \newblock {\bf 1}, 325, (1924).

\bibitem{Che74}
S.~Chern and J.~Simons.
  Ann. Math. \newblock {\bf 99}, 48, (1974).

\bibitem{Nie82}
H.~T. Nieh and M.~L. Yan.
  J. Math. Phys. \newblock {\bf 23}, 373, (1982).

\bibitem{Mie01}
E.~W. Mielke.
 Int. J. Theor. Phys. \newblock {\bf 40}, 171, (2001).

\bibitem{Gho00}
P.~Ghosh and P.~Bandyopadhyay.
 Int. J. Mod. Phys. A {\bf 15}, 3287, (2000).

\bibitem{Cha97}
O.~Chandia and J.~Zanelli.
 Phys. Rev. D {\bf 55}, 7580, (1997).
 
 \bibitem{note1}
If we
name three terms of r.h.s.  of (\ref{eqn:xb}) as
$\mathcal{L}\mbox{${}_{01}$}, \mathcal{L}\mbox{${}_{02}$}$ and $\mathcal{L}\mbox{${}_{03}$}$, then
we see that, for arbitrary values of $a,b$ and $c$,
 $a\mathcal{L}\mbox{${}_{01}+b$}\mathcal{L}\mbox{${}_{02}+c$}\mathcal{L}\mbox{${}_{03}=(b-a)$}\mathcal{L}\mbox{${}_{02}+a\times
SO(3,1)$}$-Pontryagin density $-\frac{2c}{l^{2}} \times$ Nieh-Yan
density $=(b-a)\mathcal{L}\mbox{${}_{02}$}+$ Surface  Terms(S.T.). Now, with
out any loss of generality, we assume $(b-a)=6$ and then we get
$a\mathcal{L}\mbox{${}_{01}+b$}\mathcal{L}\mbox{${}_{02}+c$}\mathcal{L}\mbox{${}_{03}$}\equiv
\mathcal{L}\mbox{${}_{0}$}+$ S.T.. By considering that, $a$  and $b$ have
unequal values, here we are assuming that the $SO(4,1)$ invariance
of the theory is broken in a $SO(3,1)$ invariant way in the real
physical world.

\bibitem{Sha01}
I.~L. Shapiro.
 Phys. Rep. {\bf 357}, 113, (2001).

\bibitem{Heh95}
{F. W. Hehl, J. D. McCrea, E. W. Mielke} and Y.~Ne'eman.
 Phys. Rep. \newblock {\bf 258}, 1, (1995).

\bibitem{note2}
An important advantage of this part of the Lagrangian is that - it is a
   quadratic one with respect to the field derivatives and this
   could be valuable in relation to the quantization program of gravity like other gauge theories of
   QFT.


\bibitem{Con94}
A.~Connes.
\newblock {\em Noncommutative Geometry}.
 Academic Press, New York, (1994).

\bibitem{Ban91}
P.~Bandyopadhyay.
\newblock {\em Quantum Space-time, Quantum Gravity and Torsion in Modern
  Problems of Theoretical Physics, Festschrift Volume Devoted to the 85th
  Jubilee of Professor D. D. Ivanenko( Eds. P. I. Pronin and Yu. N. Obukov)},
  World Scientific, Singapore, (1991).

\bibitem{Mie99}
E.~W. Mielke and D.~Kreimer.
 Gen. Rel. Grav. \newblock {\bf 31}, 701, (1999).

\bibitem{Mie99a}
E.~W. Mielke and F.~Mac\'{\i}as.
 Annals. Phys. \newblock {\bf 8}, 301, (1999).

\bibitem{Kre01}
D.~Kreimer and E.~W. Mielke.
  Phys. Rev. D \newblock {\bf 63}, 048501, (2001).

\bibitem{Obu97}
{Y. Obukhov, E. Mielke, J. Budczies} and F.~W. Hehl.
\newblock {\em On the Chiral Anomaly in Non-Riemannian Spacetimes, Biedenharn
  memorial volume, K$\ddot{\mbox{o}}$ln report (January- 1997)}, ({\tt
  gr-qc/9702025}).

\bibitem{Cha98a}
O.~Chandia and J.~Zanelli.
  Phys. Rev. D \newblock {\bf 58}, 045014, (1998).

\bibitem{Mie03}
E.~W. Mielke and A.~A.~Rinc$\acute{\mbox{o}}$n Maggiolo.
  Phys. Rev. D \newblock {\bf 68}, 104026, (2003).

\bibitem{Fil03}
A.~V. Filippenko.
\newblock {\em in Carnegi Observatories Astrophysics Series, vol-2: Measuring
  and Modeling the Universe, proceedings of a meeting held in November 2002
  (ed. W. L. Freedman )}, Cambridge University Press, Cambridge, in press, ({\tt
  astro-ph/0307139}).
  
 \bibitem{note3}
i.e., using four dimensional
stereographic coordinates.
 

\bibitem{Pad03}
T.~Padmanabhan.
  Phys. Rept. \newblock {\bf 380}, 235, (2003).

\bibitem{Mah02}
P.~Mahato.
  Mod. Phys. Lett. A \newblock {\bf 17}, 475, (2002).
  
\bibitem{Ban95}
L.~Mullick and P.~Bandyopadhyay.
 J. Math. Phys. \newblock {\bf 36},  370, (1995).
 
 
\end{thebibliography}
 \end{document}